\documentstyle[onecolumn]{mn}

\title [Star/Galaxy Separation]{ Star/Galaxy Separation Revisited : Into the Zone of Avoidance}

\author[Naim]{Avi Naim \\ \\
Dept. of Physics, Carnegie-Mellon University, 5000 Forbes Ave., Pittsburgh, 
PA 15213}

\begin{document}

\maketitle
 
\begin{abstract}

The problem of automated separation of stars and galaxies on photographic
plates is revisited with two goals in mind : First, to separate galaxies 
from everything else (as opposed to most previous work, in which galaxies 
were lumped together with all other non-stellar images). And second, to 
search optically for galaxies at low Galactic latitudes (an area that has
been largely avoided in the past). This paper demonstrates how an artificial 
neural network can be trained to achieve both goals on Schmidt plates of 
the Digitised Sky Survey. Here I present the method while its application 
to large numbers of plates is deferred to a later paper. Analysis is also 
provided of the way in which the network operates and the results are used 
to counter claims that it is a complicated and incomprehensible tool.

\end{abstract}

\section{Introduction}

The separation of stars from galaxies on photographic plates or CCDs is 
required in large surveys like the APM survey (Maddox {\it et al.} 1990a) and 
the Sloan Digital Sky survey (Gunn 1995), for purposes such as the preparation 
of target lists of galaxies for observations (e.g., in order to obtain 
redshifts). The sheer number of detected objects on survey plates forces one 
to use automated procedures for this task and in every large survey some 
such method is chosen. It is generally relatively easy to separate single 
images of stars from everything else, and different methods have been 
successfully tried in order to achieve this goal. The DAOFIND package (e.g., 
as incorporated in IRAF) fits point spread functions to detect stars. Sebok 
(1979) uses surface brightness measurements from the image and matches them to 
templates of stars and various galaxy models. Jarvis and Tyson (1981) define 
shape parameters on the basis of image moments. Slezak {\it et al.} (1988) 
combine light profile with shape parameters, while the APM survey team 
(Maddox {\it et al.} 1990a) use automated parameters derived from the APM 
microdensitometer. 

Regardless of which parameters are measured, the problem is usually formulated 
in terms of finding an optimal decision surface in the space spanned by the 
chosen parameters. This approach implies supervised learning by the classifier,
i.e., learning from examples that were pre-classified, e.g., by eye. An 
alternative approach was recently proposed by M\"{a}h\"{o}nen \& Hakala (1995),
who use an unsupervised learning method (self organising maps) to separate 
images on the basis of their appearance without classifying them by eye first. 
While promising, their approach was only demonstrated on synthetic images and 
is yet to be applied to real images. In this paper supervised learning was 
preferred. The typical output of a classifier is the drawing of linear 
boundaries which distinguish between stars and non-stellar images, and it is 
usually assumed that the vast majority of all non-stellar objects in the field 
are galaxies. 

There are at least three points that require further consideration when 
constructing yet another star/galaxy classifier :

\begin{enumerate}
\item{Which parameters are adequate for describing galaxies and stars ?}
\item{Is linear classification good enough ?}
\item{Is it true that non-stellar objects are mostly galaxies ?}
\end{enumerate}

\bigskip

Although important, the first question is difficult to answer. Different 
parameters are measured by different researchers who do not all use the same 
plate material or even the same classification tools. Since the end result is 
a combination of all of these factors it is difficult to examine the effect of 
parameter choice separately. Because of this difficulty this paper makes no 
attempt to compare the parameters used here with those chosen by others. 
Rather, I augment parameters which were used successfully elsewhere 
(Odewahn {\it et al.} 1992) by several special-purpose parameters.

The second question requires careful comparison of linear and non-linear 
classifiers, using the same parameters and the same plate material. In many 
cases linear separation of stellar from non-stellar objects in the selected 
parameter space appears adequate (as judged by the distribution of galaxies
and stars in plots of one parameter against another). However, in general a
non-linear classifier is capable of at least as good a classification as a 
linear classifier, and thus the choice of a non-linear classifier seems 
natural. This paper follows Odewahn {\it et al.} in selecting artificial 
neural networks as the (non-linear) classifier. 

As for the third question, it is shown below that galaxies do not dominate the 
population of non-stellar objects : the incidence of galaxies among 
non-stellar images is less than $50\%$ even at high Galactic latitudes. This 
implies that treating ``non-stars'' as ``galaxies'' is in general wrong. Among
the many non-stellar images on a typical plate one finds images of two or more 
overlapping stars which create elongated shapes; plate defects (e.g., 
scratches); meteorite trails; and combinations of the above with true galaxies.
The fraction of galaxies drops significantly as one moves to low Galactic 
latitudes, making the problem worse still.

The aim of this paper is to extend previous work in two respects : first, 
by constructing a classifier that is capable of telling galaxies from all other
objects. Discriminating between merged and extended objects is not an easy
task and is largely ignored in this field. A notable exception is the work
done by the APM team (Maddox {\it et al.} 1990a), who constructed two 
special parameters for telling merged objects apart from single objects. I 
attempt to carry this distinction further, by training the ANN to tell apart 
six different classes of object. The second respect in which previous work is 
extended is by making the classifier general enough to handle star/galaxy 
separation at low Galactic latitudes. The reason why one would wish to extend 
high latitude work into the {\it zone of avoidance} is the difficulty of using 
standard techniques to detect galaxies optically behind the disk of the Galaxy.
The potential for new discoveries in that part of the sky has been realised 
several times in the past few decades, most recently in the radio band by 
Kraan-Korteweg {\it et al.} (1994). An optical identification of galaxy 
candidates could help save a lot of time in the search for partially obscured 
galaxies. 

This paper serves to introduce the problem and the chosen method. The ultimate
goal is the production of an automated catalogue of galaxy candidates behind
the Galactic disk. The catalogue will be described in a later paper, which
will also address issues of data quality, plate to plate changes and
the overall reliability of the classifications. The data used for this first 
paper are described in \S~2. \S~3 describes the eyeball classification of the 
images. The parameters chosen to represent each image are disussed in \S~4 and
put to use in training the ANN, in \S~5. The discussion follows in \S~6.

\bigskip
\section{The Data}

The Digitised Sky Survey (DSS, Lasker {\it et al.} 1990) is a collection of 
digitised plates, taken with the UK Schmidt Telescope in Australia (southern 
sky, blue band) and with the Hale telescope at Mount Palomar (the POSS I 
survey of the northern sky, red band). The plates were digitised at a low 
resolution in the first generation of the DSS, and the information distributed 
on CD-ROM discs. A second generation of higher resolution scans is currently
in production. For the investigation of automated star/galaxy separation in 
this paper I used the DSS mk I collection of 61 discs covering the southern 
skies. The resolution of the PDS scanner used in the digitisation was $1.7$ 
arcsec  pixel$^{-1}$, so a reasonable lower limit on the size (major axis 
length) of treated images was set at $15''$. This roughly corresponds to a 
semi-major axis of $4.5$ pixels, and allows one to get some structural detail 
even from the smallest objects in the sample. The UK Schmidt plates are the 
same plates used for the APM survey, and the reader may obtain a full 
description of the quality and uniformity of the data in their papers (e.g., 
Maddox {\it et al.} 1990a; 1990b). In order to get an estimate of the 
magnitudes of the detected objects let us use a relation derived by the APM 
team (S. Maddox, private communication) :

\smallskip
\begin{math}
$$
(1)~~~~~~~~~~~~~~~  log(Area/(1'')^2) = 8.024 - 0.324~ b_j 
$$
\end{math}
\smallskip

\noindent where $b_j$ is the blue magnitude in the plate. This relation is
derived for magnitudes between 13.5 and 19. It not very tight for extended 
objects, and should therefore be used only as a rough guide to the magnitudes 
involved. At the limiting size of $15''$ a round object then has a rough 
magnitude of 17.8. 

Considerations of the memory required by a reduction program forced a limit of 
about 1~ deg$^2$ on any single patch of sky retrieved from the discs. In order 
to allow for images on the edges of any given patch the actual patches 
extracted were square, $70'$ on the side, having an overlap of $5'$ with 
patches on every side. The image itself then took up 12 Mb of memory, and 
various tables which were required for the reduction process increased the
overall size of the program to roughly 40 Mb. Initially data were extracted 
for 25 patches making up field 646 of the UK Schmidt survey (centered on RA,
DEC (1950): 13h, $-15^\circ$). These patches are situated at a high Galactic 
latitude (the center of the field is at latitude $50^\circ$), so no problems 
such as overcrowding by stars or significant extinction were anticipated. 
Figure 1 shows a square section, 512 pixels on the side, of one of these 
patches. This corresponds roughly to a section 15 arcmin on the side. There 
are two adjacent scratches (bottom center) and at least three galaxies in this 
patch. There are several ``stellar mergers'' (images of two or more stars 
overlapping each other) as well. The patch is not very crowded and there are 
no mergers of more than three objects. In contrast, figure 2 shows another 
square section, 512 pixels on the side, of one of the patches at low galactic 
latitude (to be described below). This patch contains many more objects than 
the high latitude patch. There are numerous stellar mergers in it, and it is 
difficult to find a galaxy that does not merge into one or more stars.

\bigskip

Ten more patches were extracted from the DSS. Five were taken at Galactic 
latitude of around $15^\circ$ (RA around 13 hours and DEC $-47.5^\circ$) and 
five at Galactic latitude of about $5^\circ$ (RA centered on 13 hours and DEC 
on $-57.5^\circ$). The quality of the images was expected to be poorer, due to 
Galactic extinction at these latitudes and crowding by stars. This is seen 
clearly in figure 2. Upon careful inspection it was obvious that the smaller 
images contained a lot of spurious structure, and their shapes were affected 
badly. It was very difficult to assign certain eyeball classifications for 
images smaller than $25''$. For this reason, a lower limit of $25''$ was 
imposed on the size of all images to be eyeballed. Using the relation (1) 
this implies a limiting magnitude for round objects of roughly 16.5, more than 
a magnitude brighter than the limit of the high latitude objects. 

\bigskip

\section{Eyeball Classification}

At high Galactic latitudes all 6240 objects larger than $15''$ were classified
by eye. Each object was assigned one of six possible classes:

\begin{enumerate}
\item{ Star}
\item{ Galaxy}
\item{ Noise (e.g., scratches on the plate)}
\item{ Merger of two or more stellar images}
\item{ Merger of one or more stellar image(s) and a galaxy image}
\item{ Merger of two or more galaxy images}
\end{enumerate}

A classification was forced even in the few cases where it was not entirely 
certain. An object would be assigned the class ``stellar merger'' even if it 
there was a significant difference in the apparent sizes of both stars. On the
other hand, the star-galaxy merger class was assigned only to images in which 
the star was not much smaller than the galaxy. This was done because any galaxy
image which is not severely contaminated by that of a star contains enough of 
the galaxy to allow its analysis and possibly even a detailed morphological 
classification. The distinction between a galaxy and a star-galaxy merger was 
therefore a little fuzzy, and could contribute some noise to the classification
set.

The distribution of classifications of the 6240 objects at high Galactic 
latitude is described in table 1. These objects were randomly divided between 
two sets (sets 1 and 2 hereafter), each containing 3120 images. The breakdown 
of each of these sets into classes is also shown in table 1. As expected, stars
are the most common class. There are quite a few galaxies, but even at such 
high Galactic latitudes the number of stellar mergers is quite {\it comparable 
to the number of galaxies}. If one were to distinguish only stars from 
everything else in these sets, galaxian objects would make up only around 
$54\%$ of the ``non-stellar objects'' class. 

\begin{table}
\centering
\caption{Distribution of Eyeball Classes in the Data Sets of field 646 (center
at RA,DEC (1950): 13h, -15$^\circ$; b = 50$^\circ$.)}
\halign{\hfil#\hfil &\quad\hfil#\hfil &\quad\hfil#\hfil &\quad\hfil#\hfil \cr
{\bf Class} & {\bf Set 1} & {\bf Set 2} & {\bf Total} \cr
{\bf Star} & 1560 & 1551 & 3111 \cr
{\bf Galaxy} & 650 & 652 & 1302 \cr
{\bf Noise} & 102 & 94 & 196 \cr
{\bf Stellar Merger} & 613 & 615 & 1228 \cr
{\bf Star-Galaxy Merger} & 184 & 190 & 374 \cr
{\bf Galaxy Merger} & 11 & 18 & 29 \cr}
\end{table}

\bigskip

At low Galactic latitudes all objects larger than 25'' were classified by eye.
The higher density of objects in these regions implied a very high number of 
objects, but the imposition of the higher size limit cut this number down to
1029. Again, these objects were divided to two sets of nearly equal sizes 
(hereafter sets 3 and 4). The distribution of classifications in these sets is
shown in table 2, and figure 3 shows the frequencies of eyeball classes in the 
high-latitude patches and in the low-latitude patches. The excess of merged 
objects as one goes to lower latitudes is evident. The predominant class by 
far at low latitudes is the stellar-merger class, making galaxies a rather
rare exception among non-stellar images.

\begin{table}
\centering
\caption{Distribution of Eyeball Classes in the Low Latitude Data Sets.}
\halign{\hfil#\hfil &\quad\hfil#\hfil &\quad\hfil#\hfil &\quad\hfil#\hfil \cr
{\bf Class} & {\bf Set 3} & {\bf Set 4} & {\bf Total} \cr
{\bf Star} & 52 & 60 & 112 \cr
{\bf Galaxy} & 10 & 12 & 22 \cr
{\bf Noise} & 6 & 5 & 11 \cr
{\bf Star Merger} & 383 & 384 & 767 \cr
{\bf Star-Galaxy Merger} & 61 & 54 & 115 \cr
{\bf Galaxy Merger} & 2 & 0 & 2 \cr}
\end{table}

\section{Data Reduction and Parameter Descriptions}

The reduction program used on the APM galaxy images (Naim {\it et al.} 1995)
was modified for the purpose of reducing these patches, although its principles
remained the same: the sky level was obtained and subtracted and a list of all
isolated groups of linked pixels was created. Each such group will be referred 
to as an ``object'' hereafter, although many of them contain several merged 
objects. A set of 17 parameters was measured for each of the objects. Many of 
these parameters had been defined and used successfully for star/galaxy 
separation by Odewahn {\it et al.} (1992). These are listed below:

\begin{itemize}
\item{log Area: The logarithm of the number of pixels in the object.}
\item{Peak Intensity: Highest intensity of any pixel in the object.}
\item{Average Intensity: Average intensity of all pixels in the object.}
\item{Ellipticity and Position Angle: Object ellipticity and position angle of
the major axis, calculated from the intensity-weighted second moments matrix 
of the image. Position angle not used to train the ANN.}
\item{Semi-major Axis Length: Obtained from the distance (along the major 
axis) from the center of the object to its edge.}
\item{Spike Trace: The ratio of the area implied by the extent of the object 
in the x and y directions of the picture to the area from pixel count. This 
ratio is sensitive to the existence of diffraction spikes, which extend along 
the x and y directions.}
\item{The following moment: 

\smallskip
\begin{math}
$$
(2)~~~~~~~~~~~~~~~  I_1 = \frac{\sum_i I(r_i)/r_i}{\sum_i I(r_i)}~,
$$
\end{math}
\smallskip

\noindent where $I(r_i)$ is the intensity at a given value of $r_i$, and $r_i$ 
is the semi-major axis of the elliptical ring to which point $i$ belongs. Both 
summations are over all pixels in the object.}
\item{Light Gradients: In analogy with the sampled ellipses of the APM galaxy 
images, here only four ellipses were sampled and averaged to give a crude light
profile consisting of four values. Five differences of intensities between 
pairs of ellipses were calculated as the gradients: The gradients between the 
pairs (1,2), (1,3), (1,4), (2,4) and (3,4) were used. This choice was made 
following Odewahn {\it et al.} (1992).}
\end{itemize}

In addition to the above list several further parameters were designed to 
address the issue of classifying merged objects as separate from single stars 
and galaxies. These are the new parameters:

\begin{itemize}
\item{Average Separability of Intensity Peaks: A list of local intensity 
maxima in the object was created, and a measure of how low the intensity 
dropped between any two such maxima was calculated. This measure was defined 
as $I_{min} / I_{peaks}$ where $I_{min}$ was the lowest intensity found along 
the line connecting the two peaks, and $I_{peaks}$ was the average intensity 
of the two peaks. The parameter ``average separability'' was then evaluated 
as the average of all such pairwise separation measures.}
\item{Multiplicity of Peaks: The number of separate intensity peaks located 
in the object, as indicated by the separation measures discussed above. A
threshold value for the intensity drop was defined (0.45) and peaks were 
grouped together in bunches which were connected by intensities higher than the
threshold. If the number of resulting such groups exceeded 3 it was set to 1, 
so the only possible values it could take were 1, 2 or 3. This was done because
of the very low likelihood of mergers of more than three images at high 
Galactic latitudes. An object with a larger number of peaks would almost 
certainly be a galaxy, but the actual number would then be meaningless, so it 
would be right to set it to 1.}
\item{Ratio of Central to Peak Intensity: In many cases the center of a merger
lies in between the centers of the individual objects making it, and the 
intensity at the geometrical center of the merger is much lower than the 
overall maximal intensity in the object. This ratio betrays these cases and
distinguishes them from the single object case, for which this ratio is 
normally 1.}
\item{Average Quadrant Intensity Difference, based on sampled ellipses: The 
sampled ellipses were divided to four groups making up four quadrants, the 
dividing lines being the major and minor axes. The total intensity in each of 
these quadrants was measured and all six differences between these quadrants 
were calculated. The six differences are averaged to produce this single
parameter.}
\item{Average Quadrant Area Difference, based on whole image: The area of each
quadrant was calculated, where in this case quadrants were defined over the 
entire object (rather than just the sampled ellipses) and the dividing lines 
were the x and y axes of the image (rather than its major and minor axes). This
parameter is the average of all six area-differences between pairs of 
quadrants.}
\end{itemize}

\section{Training the ANN}

\subsection{ANN Classification at High Latitudes}

For an overview of ANNs the reader is referred to Hertz, Krogh \& Palmer 
(1991). In the context of astronomical classification ANNs have been used, 
e.g., by Odewahn {\it et al.} (1992); Storrie-Lombardi {\it et al.} (1992); 
Serra-Ricart {\it et al.} (1993); Naim {\it et al.} (1995), to mention but a 
few. The ANN architecture we use here is 17:5:6 (5 hidden nodes in a single
hidden layer and six output nodes), where each output denotes a specific type 
of object and output values only roughly approximate Bayesian a-posteriori 
probabilities for class given data. The ANN was run ten times, each time 
starting with a different randomisation for the weights, training on set 1 and 
testing on set 2. Then the procedure was reversed for ten more runs, training 
on set 2 and testing on set 1. Each collection of ten runs was analysed 
separately. The output node values were averaged over the ten runs, and the 
classification of each pattern by the ANN was taken as the class for which the 
average of the ten runs gave the highest ``probability''. Table 3 shows the 
overall performance of the ANN in terms of classification matrices. Here, and 
in all subsequent classification matrices shown in this paper, eyeball classes 
appear on the vertical axis, each taking up a row in the matrix, and the 
resulting ANN classes appear on the horizontal axis, each taking up a column. 
Successes (``hits'') are recorded on the diagonal and all off-diagonal terms 
are misclassifications. Also shown in table 3 are the ``Hit Rate'' and ``False 
Alarm Rate'' for each class. The hit rate is defined as the fraction of 
successful classifications out of all patterns belonging to a given class,
e.g., the fraction of objects classified by eye as galaxies which were 
assigned the same class by the ANN:

\bigskip
\begin{math}
$$
(3)~~~~~~~~~~~~~~~  HR_i = 100 \times \frac{C_{ii}}{\sum_j C_{ij}}~,
$$
\end{math}
\bigskip

\noindent where $C$ is the classification matrix. The false alarm rate is the 
fraction of contaminating patterns, e.g., the fraction of patterns classified
by eye as non-stellar, which were misclassified by the ANN as stars, out of 
all patterns the ANN called stars:

\bigskip
\begin{math}
$$
(4)~~~~~~~~~~~~~~~  FA_j = 100 \times \frac{\sum_{i \neq j}C_{ij}}
                                                  {\sum_i C_{ij}}~.
$$
\end{math}
\bigskip

\begin{table}
\centering
\caption{Classification Results of the ANN (averaged over 10 runs) at High
Galactic Latitudes, for all Patterns in Set 2 (training on Set 1). Eyeball
Classes are recorded on the vertical axis and ANN Classes on the horizontal 
axis.}
\halign{\hfil#\hfil &\quad\hfil#\hfil &\quad\hfil#\hfil &\quad\hfil#\hfil 
&\quad\hfil#\hfil &\quad\hfil#\hfil &\quad\hfil#\hfil &\quad\hfil#\hfil 
&\quad\hfil#\hfil \cr
{\bf Class} & {\bf Star} & {\bf Galaxy} & {\bf Noise} & {\bf S+St} & 
{\bf St+G} & {\bf G+G} & {\bf \% HR} & {\bf \% FA} \cr
{\bf Star} & 1533 &   3 &   0 &  15 &   0 &   0 &  98.8 &  1.7 \cr
{\bf Galaxy} &  2 & 616 &   1 &  16 &  17 &   0 &  94.5 & 11.9 \cr
{\bf Noise} &   0 &   3 &  79 &   4 &   8 &   0 &  84.0 &  9.2 \cr
{\bf St+St} & 21 & 17 & 4 & 552 &  21 &   0 &  89.8 & 14.4 \cr
{\bf St+G} & 3 & 56 &   3 &  56 &  72 &   0 &  37.9 & 44.6 \cr
{\bf G+G} & 0 &   4 &   0 &   2 &  12 &   0 &   0.0 &  0.0 \cr
\cr}

\halign{\hfil#\hfil \cr
As above for all Patterns in Set 1 (training on Set 2). \cr}

\halign{\hfil#\hfil &\quad\hfil#\hfil &\quad\hfil#\hfil &\quad\hfil#\hfil 
&\quad\hfil#\hfil &\quad\hfil#\hfil &\quad\hfil#\hfil &\quad\hfil#\hfil 
&\quad\hfil#\hfil \cr
{\bf Class} & {\bf Star} & {\bf Galaxy} & {\bf Noise} & {\bf St+St} & 
{\bf St+G} & {\bf G+G} & {\bf \% HR} & {\bf \% FA} \cr
{\bf Star} & 1543 &   5 &   0 &  11 &   1 &   0 &  98.9 &  3.4 \cr
{\bf Galaxy} & 10 & 608 &   3 &  19 &  10 &   0 &  93.5 & 12.6 \cr
{\bf Noise} &   0 &   9 &  83 &   1 &   4 &   0 &  86.3 &  9.3 \cr
{\bf St+St} & 38 & 21 & 3 & 529 &  22 &   0 &  86.3 & 13.0 \cr
{\bf St+G} & 7 & 53 &   3 &  48 &  73 &   0 &  39.7 & 39.7 \cr
{\bf G+G} & 0 &   1 &   0 &   0 &  11 &   0 &   0.0 &  0.0 \cr}
\end{table}

The separation of stars from galaxies is almost perfect, with the very few 
mistakes being mostly cases in which the eyeballing was uncertain to begin with.
As expected, there is some mixing between stars and stellar mergers, and 
between galaxies and star-galaxy mergers. The galaxy-galaxy merger class is 
not well defined at all, probably due to the very small number of cases 
belonging to this class. From the point of view of creating a galaxy catalogue,
the most worrying numbers are the misclassified galaxies (incompleteness) and 
the contamination of galaxies by non-galaxian images. Although both of these 
numbers are small here, one must bear in mind that at lower Galactic latitudes 
the incidences of mergers is much higher and with them the contamination rate 
is expected to rise. Of the 615 stellar mergers in set 2, 17 were mistaken to 
be galaxies ($2.8\%$), and so were 21 of the 613 ($3.4\%$) of the stellar 
mergers in set 1. 

In order to check the dependence of the ANN performance on the sizes of the 
images each set was split in two, one part containing images whose major axis
was larger than $30''$ and the other containing images smaller than $30''$. 
The results for both data sets are shown in tables 4 and 5. The hit rate for 
galaxies improves in one set and decreases slightly in the other for the larger
images, and their contamination by non-galaxian images drops from $3.3\%$ to 
$1.4\%$ in set 2, and rises from $5.3\%$ to $6.6\%$ in set 1, so there is no 
overall improvement. Following Odewahn {\it et al.} (1992), the next thing to 
try is separating the images into two size groups before training, and train a 
separate ANN for each size group. The dataset was split in two, including all 
images larger than $25''$ in one group, and all images smaller than $30''$ in 
the other, so there is some overlap between the groups. As expected, there were
many more objects in the ``small'' sets than in the ``big'' sets. However, the 
results (tables 6 and 7) show that there is virtually no change in the 
performance of the ANN, and so the conclusion is that the chosen parameters 
describe the various classes of objects well regardless of image size, down to
the limit chosen to begin with.

It is interesting to compare the performance of the ANN at high Galactic 
latitudes with the result of Maddox {\it et al.} (1990a), who used empirical 
functional dependencies between various APM-measured parameters. Their sample
was defined differently, by imposing an area limit of 16 pixels, regardless of
the shape of the image. They used scans with a finer resolution, of roughly
0.5 arcsec/pixel, and the chosen size limit implies the inclusion of
fainter objects. The parameters they used were automatically measured by the
APM machine and do not allow the analysis of merged objects. For the separation
between stars and non-stars they got a hit rate of 90-95$\%$ (depending on 
magnitude) and a contamination level of 5-10$\%$. Comparing this with the ANN
hit rate of over 99$\%$, it appears that the ANN is doing much better. However,
it is hardly surprising, because of the inclusion of fainter and smaller 
objects in their sample.

\begin{table}
\centering
\caption{Classification Results of the ANN for Images larger than $30''$ in
Set 2 (training on Set 1). Eyeball Classes are recorded on the vertical axis 
and ANN Classes on the horizontal axis.}
\halign{\hfil#\hfil &\quad\hfil#\hfil &\quad\hfil#\hfil &\quad\hfil#\hfil 
&\quad\hfil#\hfil &\quad\hfil#\hfil &\quad\hfil#\hfil &\quad\hfil#\hfil 
&\quad\hfil#\hfil \cr
{\bf Class} & {\bf Star} & {\bf Galaxy} & {\bf Noise} & {\bf St+St} & 
{\bf St+G} & {\bf G+G} & {\bf \% HR} & {\bf \% FA} \cr
{\bf Star} &  243 &   0 &   0 &   2 &   0 &   0 &  99.2 &  3.2 \cr
{\bf Galaxy} &  2 & 196 &   0 &   3 &   9 &   0 &  93.3 &  7.1 \cr
{\bf Noise} &   0 &   0 &  15 &   1 &   0 &   0 &  93.8 &  0.0 \cr
{\bf St+St} & 6 & 3 &   0 &  76 &   2 &   0 &  87.4 & 16.5 \cr
{\bf St+G} & 0 & 10 &   0 &   9 &   3 &   0 &  13.6 & 82.4 \cr
{\bf G+G} & 0 &   2 &   0 &   0 &   3 &   0 &   0.0 &  0.0 \cr
\cr}
\halign{\hfil#\hfil \cr
As above for Images smaller than $30''$ in Set 2 (training on Set 1). \cr}
\halign{\hfil#\hfil &\quad\hfil#\hfil &\quad\hfil#\hfil &\quad\hfil#\hfil 
&\quad\hfil#\hfil &\quad\hfil#\hfil &\quad\hfil#\hfil &\quad\hfil#\hfil 
&\quad\hfil#\hfil \cr
{\bf Class} & {\bf Star} & {\bf Galaxy} & {\bf Noise} & {\bf St+St} & 
{\bf S+G} & {\bf G+G} & {\bf \% HR} & {\bf \% FA} \cr
{\bf Star} & 1290 &   3 &   0 &  13 &   0 &   0 &  98.8 &  1.4 \cr
{\bf Galaxy} &  0 & 420 &   1 &  13 &   8 &   0 &  95.0 & 13.9 \cr
{\bf Noise} &   0 &   3 &  64 &   3 &   8 &   0 &  82.1 & 11.1 \cr
{\bf St+St} & 15 & 14 & 4 & 476 &  19 &   0 &  90.2 & 14.1 \cr
{\bf St+G} & 3 & 46 &   3 &  47 &  69 &   0 &  41.1 & 38.9 \cr
{\bf G+G} & 0 &   2 &   0 &   2 &   9 &   0 &   0.0 &  0.0 \cr}
\end{table}

\begin{table}
\centering
\caption{As above for Images larger than $30''$ in Set 1 (training on Set 2).}
\halign{\hfil#\hfil &\quad\hfil#\hfil &\quad\hfil#\hfil &\quad\hfil#\hfil 
&\quad\hfil#\hfil &\quad\hfil#\hfil &\quad\hfil#\hfil &\quad\hfil#\hfil 
&\quad\hfil#\hfil \cr
{\bf Class} & {\bf Star} & {\bf Galaxy} & {\bf Noise} & {\bf St+St} & 
{\bf St+G} & {\bf G+G} & {\bf \% HR} & {\bf \% FA} \cr
{\bf Star} &  254 &   2 &   0 &   3 &   0 &   0 &  98.1 &  5.9 \cr
{\bf Galaxy} &  0 & 183 &   2 &   0 &   5 &   0 &  96.3 & 13.7 \cr
{\bf Noise} &   0 &   4 &  18 &   0 &   0 &   0 &  81.8 & 18.2 \cr
{\bf St+St} & 15 & 8 &  0 &  70 &   4 &   0 &  72.2 & 12.5 \cr
{\bf St+G} & 1 & 15 &   2 &   7 &  11 &   0 &  30.6 & 50.0 \cr
{\bf G+G} & 0 &   0 &   0 &   0 &   2 &   0 &   0.0 &  0.0 \cr
\cr}
\halign{\hfil#\hfil \cr
As above for Images smaller than $30''$ in Set 1 (Training on Set 2). \cr}
\halign{\hfil#\hfil &\quad\hfil#\hfil &\quad\hfil#\hfil &\quad\hfil#\hfil 
&\quad\hfil#\hfil &\quad\hfil#\hfil &\quad\hfil#\hfil &\quad\hfil#\hfil 
&\quad\hfil#\hfil \cr
{\bf Class} & {\bf Star} & {\bf Galaxy} & {\bf Noise} & {\bf St+St} & 
{\bf St+G} & {\bf G+G} & {\bf \% HR} & {\bf \% FA} \cr
{\bf Star} & 1299 &   3 &   0 &   8 &   1 &   0 &  99.1 &  2.9 \cr
{\bf Galaxy} & 10 & 425 &   1 &  19 &   5 &   0 &  92.4 & 12.2 \cr
{\bf Noise} &   0 &   5 &  70 &   1 &   4 &   0 &  87.5 &  6.7 \cr
{\bf St+St} & 23 & 13 & 3 & 459 &  18 &   0 &  89.0 & 13.1 \cr
{\bf St+G} & 6 & 38 &   1 &  41 &  62 &   0 &  41.9 & 37.4 \cr
{\bf G+G} & 0 &   0 &   0 &   0 &   9 &   0 &   0.0 &  0.0 \cr}
\end{table}

\begin{table}
\centering
\caption{Classification Results of the ANN for Big Images Datasets; Trained on 
Set 1 and Tested on Set 2.}
\halign{\hfil#\hfil &\quad\hfil#\hfil &\quad\hfil#\hfil &\quad\hfil#\hfil 
&\quad\hfil#\hfil &\quad\hfil#\hfil &\quad\hfil#\hfil &\quad\hfil#\hfil 
&\quad\hfil#\hfil \cr
{\bf Class} & {\bf Star} & {\bf Galaxy} & {\bf Noise} & {\bf St+St} & 
{\bf St+G} & {\bf G+G} & {\bf \% HR} & {\bf \% FA} \cr
{\bf Star} &  303 &   1 &   0 &   2 &   0 &   0 &  99.0 &  6.2 \cr
{\bf Galaxy} &  0 & 231 &   3 &   3 &   5 &   0 &  95.5 &  8.0 \cr
{\bf Noise} &   0 &   3 &  16 &   0 &   0 &   0 &  84.2 & 20.0 \cr
{\bf St+St} & 17 & 2 &  0 &  96 &   3 &   0 &  81.4 & 18.6 \cr
{\bf St+G} & 3 & 11 &   1 &  14 &   9 &   0 &  23.7 & 47.1 \cr
{\bf G+G} & 0 &   3 &   0 &   3 &   0 &   0 &   0.0 &  0.0 \cr
\cr}
\halign{\hfil#\hfil \cr
As above, Trained on Set 2 and Tested on Set 1. \cr}
\halign{\hfil#\hfil &\quad\hfil#\hfil &\quad\hfil#\hfil &\quad\hfil#\hfil 
&\quad\hfil#\hfil &\quad\hfil#\hfil &\quad\hfil#\hfil &\quad\hfil#\hfil 
&\quad\hfil#\hfil \cr
{\bf Class} & {\bf Star} & {\bf Galaxy} & {\bf Noise} & {\bf St+St} & 
{\bf St+G} & {\bf G+G} & {\bf \% HR} & {\bf \% FA} \cr
{\bf Star} &  303 &   1 &   0 &  13 &   0 &   0 &  95.6 &  2.6 \cr
{\bf Galaxy} &  1 & 220 &   1 &   1 &   2 &   0 &  97.8 &  9.5 \cr
{\bf Noise} &   0 &   2 &  26 &   0 &   0 &   0 &  92.9 & 10.3 \cr
{\bf St+St} &  7 & 2 &  0 & 108 &   1 &   0 &  91.5 & 18.2 \cr
{\bf St+G} & 0 & 17 &   2 &   9 &   7 &   0 &  20.0 & 50.0 \cr
{\bf G+G} & 0 &   1 &   0 &   1 &   4 &   0 &   0.0 &  0.0 \cr}
\end{table}

\begin{table}
\centering
\caption{Classification Results of the ANN for Small Images Dataset; Trained on
Set 1 and Tested on Set 2.} 
\halign{\hfil#\hfil &\quad\hfil#\hfil &\quad\hfil#\hfil &\quad\hfil#\hfil 
&\quad\hfil#\hfil &\quad\hfil#\hfil &\quad\hfil#\hfil &\quad\hfil#\hfil 
&\quad\hfil#\hfil \cr
{\bf Class} & {\bf Star} & {\bf Galaxy} & {\bf Noise} & {\bf St+St} & 
{\bf St+G} & {\bf G+G} & {\bf \% HR} & {\bf \% FA} \cr
{\bf Star} & 1332 &   3 &   0 &  13 &   2 &   0 &  98.7 &  2.1 \cr
{\bf Galaxy} &  7 & 466 &   3 &  14 &   9 &   0 &  93.4 & 14.0 \cr
{\bf Noise} &   0 &   5 &  83 &   2 &   6 &   0 &  86.5 & 15.3 \cr
{\bf St+St} & 19 & 20 & 7 & 509 &  21 &   0 &  88.4 & 12.8 \cr
{\bf St+G} & 3 & 47 &   5 &  45 &  72 &   0 &  41.9 & 39.5 \cr
{\bf G+G} & 0 &   1 &   0 &   1 &   9 &   0 &   0.0 &  0.0 \cr
\cr}
\halign{\hfil#\hfil \cr
As above, Trained on Set 2 and Tested on Set 1. \cr}
\halign{\hfil#\hfil &\quad\hfil#\hfil &\quad\hfil#\hfil &\quad\hfil#\hfil 
&\quad\hfil#\hfil &\quad\hfil#\hfil &\quad\hfil#\hfil &\quad\hfil#\hfil 
&\quad\hfil#\hfil \cr
{\bf Class} & {\bf Star} & {\bf Galaxy} & {\bf Noise} & {\bf St+St} & 
{\bf St+G} & {\bf G+G} & {\bf \% HR} & {\bf \% FA} \cr
{\bf Star} & 1410 &   2 &   0 &  14 &   0 &   0 &  98.9 &  1.7 \cr
{\bf Galaxy} &  4 & 464 &   0 &  13 &   7 &   0 &  95.1 & 11.6 \cr
{\bf Noise} &   0 &   6 &  68 &   1 &   5 &   0 &  85.0 &  8.1 \cr
{\bf St+St} & 15 &  9 & 1 & 487 &  24 &   0 &  90.9 & 14.4 \cr
{\bf St+G} & 6 & 42 &   4 &  54 &  55 &   0 &  34.2 & 45.0 \cr
{\bf G+G} & 0 &   2 &   1 &   0 &   9 &   0 &   0.0 &  0.0 \cr}
\end{table}

\bigskip

\subsection{Extension to Lower Latitudes}

Automated classifications of the images at low Galactic latitudes were first
done using the ANNs trained on the high-latitude images, in order to see 
whether the parameters measured at high latitudes describe the distinct classes
as well in low latitudes. The results are shown in table 8.

\begin{table}
\centering
\caption{Classification Results for All Objects of Low Galactic Latitudes. The 
ANN was Trained on Set 2 of the High Galactic Latitude Data.}
\halign{\hfil#\hfil &\quad\hfil#\hfil &\quad\hfil#\hfil &\quad\hfil#\hfil 
&\quad\hfil#\hfil &\quad\hfil#\hfil &\quad\hfil#\hfil &\quad\hfil#\hfil 
&\quad\hfil#\hfil \cr
{\bf Class} & {\bf Star} & {\bf Galaxy} & {\bf Noise} & {\bf St+St} & 
{\bf St+G} & {\bf G+G} & {\bf \% HR} & {\bf \% FA} \cr
{\bf Star} &   97 &   2 &   0 &  12 &   1 &   0 &  86.6 & 71.0 \cr
{\bf Galaxy} &  1 &  18 &   0 &   1 &   2 &   0 &  81.8 & 78.3 \cr
{\bf Noise} &   0 &   0 &   9 &   1 &   1 &   0 &  81.8 & 10.0 \cr
{\bf St+St} & 229 & 49 & 0 & 450 & 39 &   0 &  58.7 & 16.4 \cr
{\bf St+G} & 7 & 13 &   1 &  74 &  20 &   0 &  17.4 & 68.8 \cr
{\bf G+G} & 0 &   1 &   0 &   0 &   1 &   0 &   0.0 &  0.0 \cr
\cr}
\halign{\hfil#\hfil \cr
As above, with the ANN Trained on Set 1 of the High Galactic Latitude 
Data. \cr}
\halign{\hfil#\hfil &\quad\hfil#\hfil &\quad\hfil#\hfil &\quad\hfil#\hfil 
&\quad\hfil#\hfil &\quad\hfil#\hfil &\quad\hfil#\hfil &\quad\hfil#\hfil 
&\quad\hfil#\hfil \cr
{\bf Class} & {\bf Star} & {\bf Galaxy} & {\bf Noise} & {\bf St+St} & 
{\bf St+G} & {\bf G+G} & {\bf \% HR} & {\bf \% FA} \cr
{\bf Star} &   97 &   3 &   1 &  10 &   1 &   0 &  86.6 & 67.6 \cr
{\bf Galaxy} &  0 &  21 &   0 &   0 &   1 &   0 &  95.5 & 85.2 \cr
{\bf Noise} &   0 &   2 &   4 &   1 &   4 &   0 &  36.4 & 81.0 \cr
{\bf St+St} &199 & 87 &15 & 434 &  32 &   0 &  56.6 & 16.2 \cr
{\bf St+G} & 3 & 28 &   1 &  73 &  10 &   0 &   8.7 & 79.6 \cr
{\bf G+G} & 0 &   1 &   0 &   0 &   1 &   0 &   0.0 &  0.0 \cr}
\end{table}

It is evident that the success rates are lower for all classes. While the 
hit rates for stars and galaxies are still rather good, the contamination of 
both by stellar mergers is huge. One factor which could contribute to this 
high contamination rate is obvious from figure 3, namely the fact that class 
priors are markedly different between the high Galactic latitude patches and 
those at low Galactic latitudes. Since the ANN roughly approximates Bayesian 
a-posteriori probabilities when it finishes training, it should in principle 
be easy to correct for this factor by dividing all probabilities by the class 
priors at high latitudes and multiplying them instead by the priors 
corresponding to the low latitudes. This was done next, and the results are 
shown in table 9. As can be seen, this indeed improves the results, but they 
are still not nearly as good as those for the high latitudes. Another possible
contributor to the lower success rates is the difference in appearance of 
objects at high and low latitudes. The fact that correcting the priors did
not solve the problem completely means that this factor was important as well. 
The conclusion was that a separate ANN should be trained on the low latitude 
data. 

A second ANN was then trained on half the low latitude data and tested on the 
other half (10 times over, as before), and then the order was reversed. Results
are shown in table 10. The hit rates for stars and galaxies are still not 
very high this time, but the false alarm rates have gone very low, especially 
for galaxies. The main errors with identifying stars is in separating them from
stellar mergers, not from other classes. While galaxies are clearly incomplete,
they are only little contaminated by stellar mergers. Of course, the number
of galaxies is small, and so even one contaminating stellar merger would raise
the contamination level significantly. Nevertheless, there are many stellar
mergers and none of them contaminates the galaxy class.

\begin{table}
\centering
\caption{Classification Results for All Objects of Low Galactic Latitudes. The 
ANN was Trained on Set 1 of the High Galactic Latitude Data and the Priors 
were Corrected to Represent the Low Latitude Distribution of Classes.}
\halign{\hfil#\hfil &\quad\hfil#\hfil &\quad\hfil#\hfil &\quad\hfil#\hfil 
&\quad\hfil#\hfil &\quad\hfil#\hfil &\quad\hfil#\hfil &\quad\hfil#\hfil 
&\quad\hfil#\hfil \cr
{\bf Class} & {\bf Star} & {\bf Galaxy} & {\bf Noise} & {\bf St+St} & 
{\bf St+G} & {\bf G+G} & {\bf \% HR} & {\bf \% FA} \cr
{\bf Star} &   96 &   0 &   0 &  15 &   1 &   0 &  85.7 & 68.0 \cr
{\bf Galaxy} &  1 &  10 &   0 &   7 &   4 &   0 &  45.5 & 61.5 \cr
{\bf Noise} &   0 &   0 &   5 &   4 &   2 &   0 &  45.5 &  0.0 \cr
{\bf St+St} & 199 & 12 & 0 & 525 & 31 &   0 &  68.4 & 17.6 \cr
{\bf St+G} &  4 & 3 &   0 &  86 &  22 &   0 &  19.1 & 63.9 \cr
{\bf G+G} & 0 &   1 &   0 &   0 &   1 &   0 &   0.0 &  0.0 \cr
\cr}
\halign{\hfil#\hfil \cr
As above, the ANN Trained on Set 2 of the High Galactic Latitude Data. \cr}
\halign{\hfil#\hfil &\quad\hfil#\hfil &\quad\hfil#\hfil &\quad\hfil#\hfil 
&\quad\hfil#\hfil &\quad\hfil#\hfil &\quad\hfil#\hfil &\quad\hfil#\hfil 
&\quad\hfil#\hfil \cr
{\bf Class} & {\bf Star} & {\bf Galaxy} & {\bf Noise} & {\bf St+St} & 
{\bf St+G} & {\bf G+G} & {\bf \% HR} & {\bf \% FA} \cr
{\bf Star} &   95 &   2 &   1 &  14 &   0 &   0 &  84.8 & 64.9 \cr
{\bf Galaxy} &  1 &  14 &   0 &   4 &   3 &   0 &  63.6 & 70.8 \cr
{\bf Noise} &   0 &   0 &   5 &   4 &   2 &   0 &  45.5 & 37.5 \cr
{\bf St+St} &173 & 24 & 2 & 543 &  25 &   0 &  70.8 & 16.3 \cr
{\bf St+G} & 2 &  7 &   0 &  84 &  22 &   0 &  19.1 & 58.5 \cr
{\bf G+G} & 0 &   1 &   0 &   0 &   1 &   0 &   0.0 &  0.0 \cr}
\end{table}

\begin{table}
\centering
\caption{Classification Results at Low Galactic Latitudes. The ANN was Trained 
and tested on Low Galactic Latitude Data, Training on Set 3 and Testing on Set 
4.}
\halign{\hfil#\hfil &\quad\hfil#\hfil &\quad\hfil#\hfil &\quad\hfil#\hfil 
&\quad\hfil#\hfil &\quad\hfil#\hfil &\quad\hfil#\hfil &\quad\hfil#\hfil 
&\quad\hfil#\hfil \cr
{\bf Class} & {\bf Star} & {\bf Galaxy} & {\bf Noise} & {\bf St+St} & 
{\bf St+G} & {\bf G+G} & {\bf \% HR} & {\bf \% FA} \cr
{\bf Star} &   28 &   0 &   0 &  32 &   0 &   0 &  46.7 & 28.2 \cr
{\bf Galaxy} &  0 &   4 &   0 &   4 &   4 &   0 &  33.3 &  0.0 \cr
{\bf Noise} &   0 &   0 &   3 &   2 &   0 &   0 &  60.0 &  0.0 \cr
{\bf St+St} & 11 &  0 & 0 & 371 &   2 &   0 &  96.6 & 19.0 \cr
{\bf St+G} & 0 &  0 &   0 &  49 &   5 &   0 &   9.3 & 54.5 \cr
{\bf G+G} & 0 &   0 &   0 &   0 &   0 &   0 &   0.0 &  0.0 \cr
\cr}
\halign{\hfil#\hfil \cr
As above, the ANN Trained on Set 2 of the Low Galactic Latitude Data.  \cr}
\halign{\hfil#\hfil &\quad\hfil#\hfil &\quad\hfil#\hfil &\quad\hfil#\hfil 
&\quad\hfil#\hfil &\quad\hfil#\hfil &\quad\hfil#\hfil &\quad\hfil#\hfil 
&\quad\hfil#\hfil \cr
{\bf Class} & {\bf Star} & {\bf Galaxy} & {\bf Noise} & {\bf St+St} & 
{\bf St+G} & {\bf G+G} & {\bf \% HR} & {\bf \% FA} \cr
{\bf Star} &   19 &   0 &   0 &  33 &   0 &   0 &  36.5 & 40.6 \cr
{\bf Galaxy} &  0 &   4 &   0 &   4 &   2 &   0 &  40.0 &  0.0 \cr
{\bf Noise} &   0 &   0 &   3 &   2 &   1 &   0 &  50.0 & 25.0 \cr
{\bf St+St} & 13 &  0 & 1 & 366 &   3 &   0 &  95.6 & 21.5 \cr
{\bf St+G} & 0 &  0 &   0 &  59 &   2 &   0 &   3.3 & 75.0 \cr
{\bf G+G} & 0 &   0 &   0 &   2 &   0 &   0 &   0.0 &  0.0 \cr}
\end{table}

\subsection{Using The ``Probabilities''}

Up to this point the only use of the actual ``probabilities'' was to determine
which is the most likely class for any given object. Given a sample of objects
with no eyeball classes one could indeed look no further than the most probable
class for each of them. However, examining only the winning class will result 
in incompleteness, which could be a serious problem. It is interesting to see 
whether the incompleteness could be reduced by examining the probability of 
belonging to other classes as well. In table 11 each entry represents, as 
before, a combination of a given eyeball class (row) and an ANN class (column).
The actual value of each entry is the probability of the ANN assigning the 
eyeball class, averaged over all cases corresponding to that given combination 
of eyeball/ANN classes. To make this clearer, examine the second row in the
upper table, corresponding to eyeball class ``galaxy'': The entry corresponding
to ANN class ``galaxy'' has the value 0.43, which is the average value of the 
``galaxy'' output node when the ANN succeeds in classifying eyeball galaxies 
correctly. For comparison, eyeball galaxies that are misclassified by the ANN 
as either stellar mergers or star-galaxy mergers average 0.14 and 0.25, 
respectively, for the ``galaxy'' output node. The fact that the former of these
numbers is smaller than the latter implies that it might be worth considering
as galaxy candidates all the objects for which the ``galaxy'' output node is 
larger than 0.20 or so, even if the resulting ANN class is not ``galaxy''. The 
reason is that even for misclassifications values larger than 0.20 are more
likely to happen for mergers involving a galaxy than for ``pure'' stellar
mergers. Of course, this will mean a rise in the level of contamination for
galaxies, but the success rate will benefit.

This analysis allows for more candidates of a given type, potentially reducing
the incompleteness, at the possible expense of raising the level of
contamination. Where one draws the line depends on where one would like to
put the interplay between completeness and contamination. As can be seen in the
lower part of table 11, which refers to the second set of runs, the picture is 
essentially the same but the actual numbers are slightly different. This 
means that caution ought to be exercised when classifying fresh data on the 
basis of a given ANN run: One should always make the decision on where to draw 
the probabilistic lines on the basis of the average probability table 
corresponding to that particular run.

\begin{table}
\centering
\caption{Averaged Output Values at Low Galactic Latitudes. The ANN was Trained 
on Set 1 of the Low Galactic Latitude Data and Tested on Set 2 of the Low
Latitude Data.}
\halign{\hfil#\hfil &\quad\hfil#\hfil &\quad\hfil#\hfil &\quad\hfil#\hfil 
&\quad\hfil#\hfil &\quad\hfil#\hfil &\quad\hfil#\hfil \cr
{\bf Class} & {\bf Star} & {\bf Galaxy} & {\bf Noise} & {\bf St+St} & 
{\bf St+G} & {\bf G+G} \cr
{\bf Star} &    0.56 &    0 &   0 &0.20 &   0 &   0 \cr
{\bf Galaxy} &     0 & 0.43 &   0 &0.14 &0.25 &   0 \cr
{\bf Noise} &      0 &    0 &0.41 &0.31 &   0 &   0 \cr
{\bf St+St} &0.38&    0 &   0 &0.75 &0.28 &   0 \cr
{\bf St+G} &   0 &    0 &   0 &0.17 &0.46 &   0 \cr
{\bf G+G} &    0 &    0 &   0 &   0 &   0 &   0 \cr
\cr}
\halign{\hfil#\hfil \cr
As above, the ANN Trained on Set 2 of the Low Galactic Latitude Data.  \cr}
\halign{\hfil#\hfil &\quad\hfil#\hfil &\quad\hfil#\hfil &\quad\hfil#\hfil 
&\quad\hfil#\hfil &\quad\hfil#\hfil &\quad\hfil#\hfil \cr
{\bf Class} & {\bf Star} & {\bf Galaxy} & {\bf Noise} & {\bf St+St} & 
{\bf St+G} & {\bf G+G} \cr
{\bf Star} &    0.62 &    0 &   0 &0.22 &   0 &   0 \cr
{\bf Galaxy} &     0 & 0.51 &   0 &0.08 &0.18 &   0 \cr
{\bf Noise} &      0 &    0 &0.39 &0.15 &0.25 &   0 \cr
{\bf St+St} &0.35&    0 &0.29 &0.76 &0.33 &   0 \cr
{\bf St+G} &   0 &    0 &   0 &0.18 &0.42 &   0 \cr
{\bf G+G} &    0 &    0 &   0 &   0 &   0 &   0 \cr}
\end{table}

\subsection{Analysis of ANN Weights: Specialisation}

It has been claimed in the past that ANNs are objectionable because they are
``black boxes'' that can not be understood. In fact, mathematically ANNs are
much simpler than common tools of the astronomical trade, such as SPH codes 
for numerical simulations, and their resulting weights can be readily 
interpreted. As described above, ten different ANNs were trained, starting each
with a different set of random weights. In general there may be many more than 
one set of final weights giving roughly the same minimised error, since the 
problem has a very large number of degrees of freedom (these ANNs were using a 
17:5:6 architecture, with 126 weights each). So, rather than trying to fit the 
weights from one run to those from another, single runs were examined closely. 
They were all found to give essentially the same results. Figures 4 and 5 show 
the weights at the end of one such run. Figure 4 depicts the weights between 
the input nodes and the hidden nodes, while figure 5 shows the weights 
connecting the hidden nodes to the outputs. In figure 5 it can be seen that 
each output performs a unique transformation of the hidden nodes connected to 
it: The weights connected to each output node ``specialise'' in picking out 
the class represented by that node. For example, the most significant 
contributions to the output node representing stars come from hidden nodes 1 
and 2, both with a positive sign. On the other hand, the weights connected to 
the output node denoting galaxies are most significant for hidden nodes 2 
(negative), 4 (negative) and 5 (positive). It also clear that the weights 
connected to the output node denoting galaxy/galaxy mergers are all very small,
which explains why this node is never chosen by the ANN in the post-training 
classification stage. One can now take this general picture one 
step back, to try and see which inputs contribute most to each of the hidden 
nodes. This could tell which input parameters are more important than 
others, for each class. In figure 4 it can be seen, for example, that hidden 
node 1 is most influenced by inputs 4 and 5, corresponding to ellipticity and 
semi-major axis length. Hidden node 2 is most influenced by inputs 6 and 15, 
which correspond to the the spike-trace and to the ratio of central-to-peak 
intensity. The largest contributions to hidden node 4 are from inputs 1, 2, 7, 
9 and 12, corresponding to log(area), peak intensity, first moment, and light 
gradients 1-3 and 2-4. Hidden node 5 is mostly dominated by inputs 2, 4, 5 and 
6, which are peak intensity, ellipticity, semi-major axis length and the 
spike-trace. 

By nature, the ANN mixes parameters together to find optimal non-linear 
combinations. It is therefore not to be expected that a ``clean'' picture of 
which parameters determine which class should emerge. However, this analysis 
can help in locating parameters which contribute almost nothing to the 
classifications, and tell us which parameters are the most important. The 
picture obviously becomes more complicated when one analyses the connections 
between the inputs and the hidden layer, but it can be safely stated that 
classes become decoupled at the hidden layer, where one can see exactly how 
they separate from each other and which ones (e.g., the galaxy-galaxy mergers 
in this case) are not well defined.

\section{Discussion}

Automated star/galaxy separation is not a new application in astronomy, but 
surprisingly work to date has left much to be desired. Separating stars from
everything else is not enough for researchers whose main interest is galaxies,
and we show that galaxies can be reliably separated from other non-stellar
images at high Galactic latitudes. However, the potential for exciting
discoveries is naturally larger in areas that were traditionally avoided, such
as fields at low Galactic latitudes. We show that it is possible to perform the
separation of galaxies from other objects in the Zone of Avoidance, although
the success rates are lower and the contamination higher.

The tool used here in the role of automated classifier is an ANN. Other 
classifiers exist, of course, and should give similar results with our
data and its parametrisation. The choice of a classifier is largely a matter of
convenience. Most of the hard work goes into the parametrisation of the 
problem. Nevertheless, ANNs have probabilistic capabilities that are useful,
e.g., in trying to improve detection limits, and they are easy to use and very
versatile. It has been claimed in the past that ANNs are slow, complicated and
are difficult to understand or interpret. The ANN code used here, which was 
kindly supplied by B. Ripley, takes few CPU minutes to converge on a 
conventional workstation and is very easy to implement. The analysis of the
ANN weights (\S~5.4 above) shows that ANNs can be understood and provide 
insight as to which parameters are more important than others.

The methods described here are readily applicable to any number of fields
extracted from the DSS. It is quite possible to proceed and catalogue the 
entire sky in this manner. However, it might be of limited interest to proceed 
with the high Galactic latitude patches, as the available digitised plates do 
not go very deep and the sky has been mapped at these latitudes and down to the
chosen limiting size. However, it is very interesting and apparently feasible 
to produce such a catalogue for the Zone of Avoidance, from which target lists 
for pointed observations of galaxies could be easily prepared. This project
will be pursued further once the higher resolution DSS mk II becomes available.

\bigskip
{\bf ACKNOWLEDGEMENTS}
\bigskip

The work described in this paper constituted one part of my Ph.D. thesis, under
the inspirational supervision of Ofer Lahav. I am indebted to him for much 
insight into ANNs and their statistical nature. I am enormously grateful to
Steve Maddox and Steve Odewahn for sharing their experience with me and
bearing with me through long discussions. I would like to thank Brian Ripley 
for allowing me to use his ANN code, and to acknowledge an Isaac Newton 
studentship which supported me through part of this research.

\newpage

\bigskip
Figure Captions :

\bigskip
Figure 1 : A 15'x15' Section of one of the Patches taken from Field
646. Compare with figure 2.

\bigskip
Figure 2 : A 15'x15' Section of one of the Low Galactic Latitude 
Patches. Compare with figure 1.

\bigskip
Figure 3 : Distributions of Eyeball-Classes in the High Latitude and Low 
Latitude Datasets. Note the total dominance of stellar mergers at low 
Galactic latitudes.

\bigskip
Figure 4 : Weights Connecting Inputs to Hidden Layer Nodes for a Single Run.

\bigskip
Figure 5 : Weights Connecting Hidden Layer Nodes to Outputs for a Single Run.

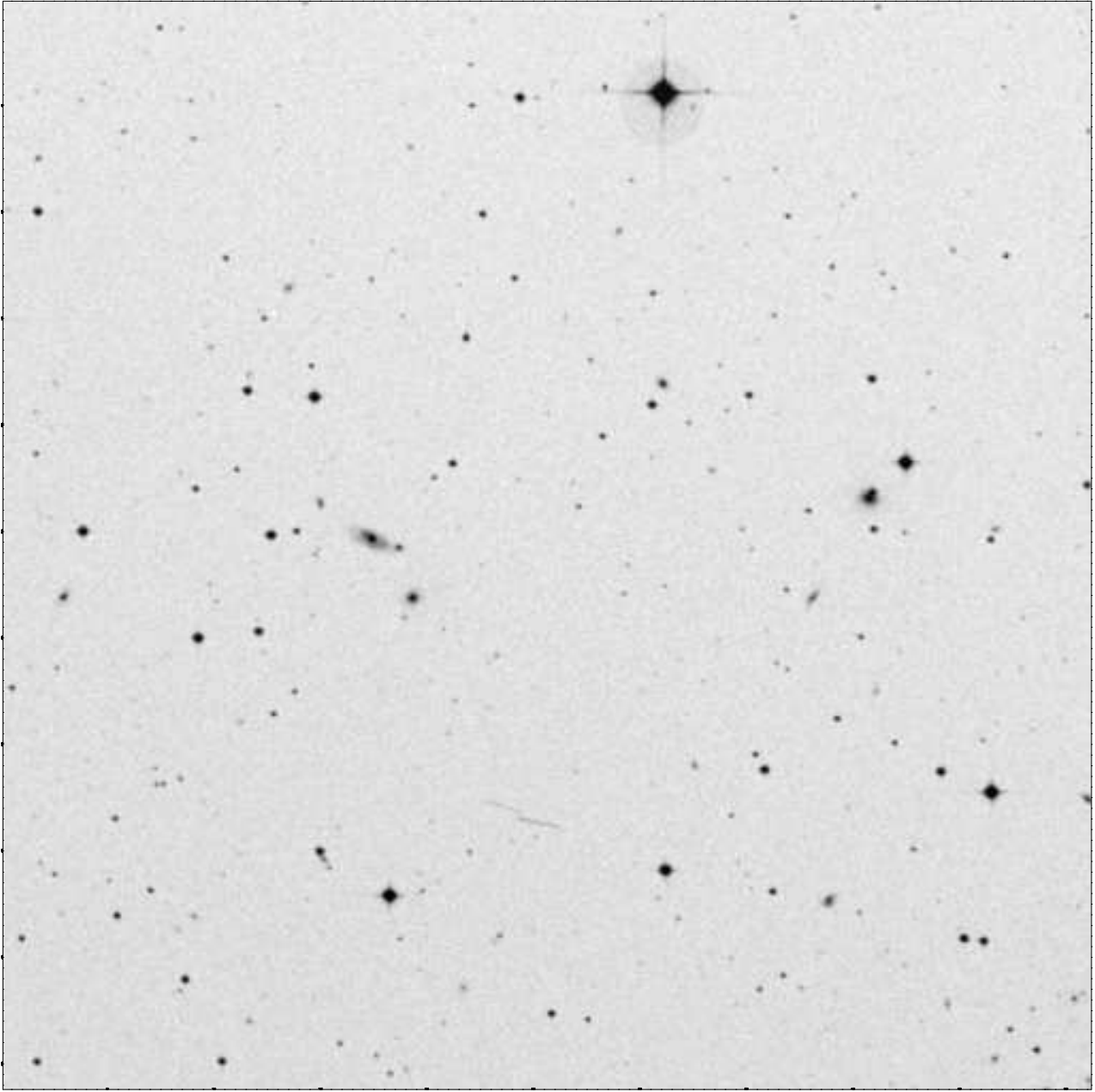
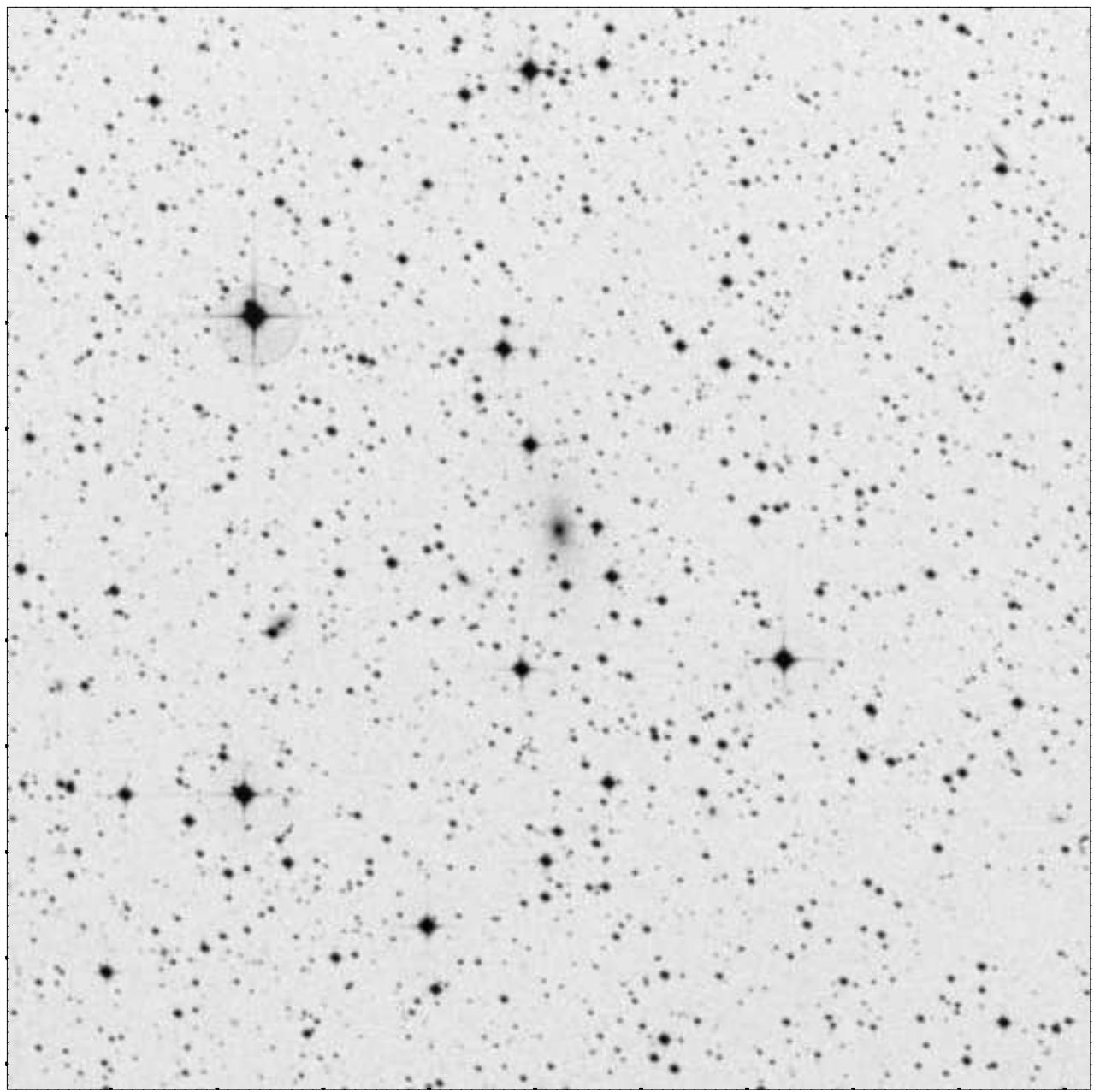
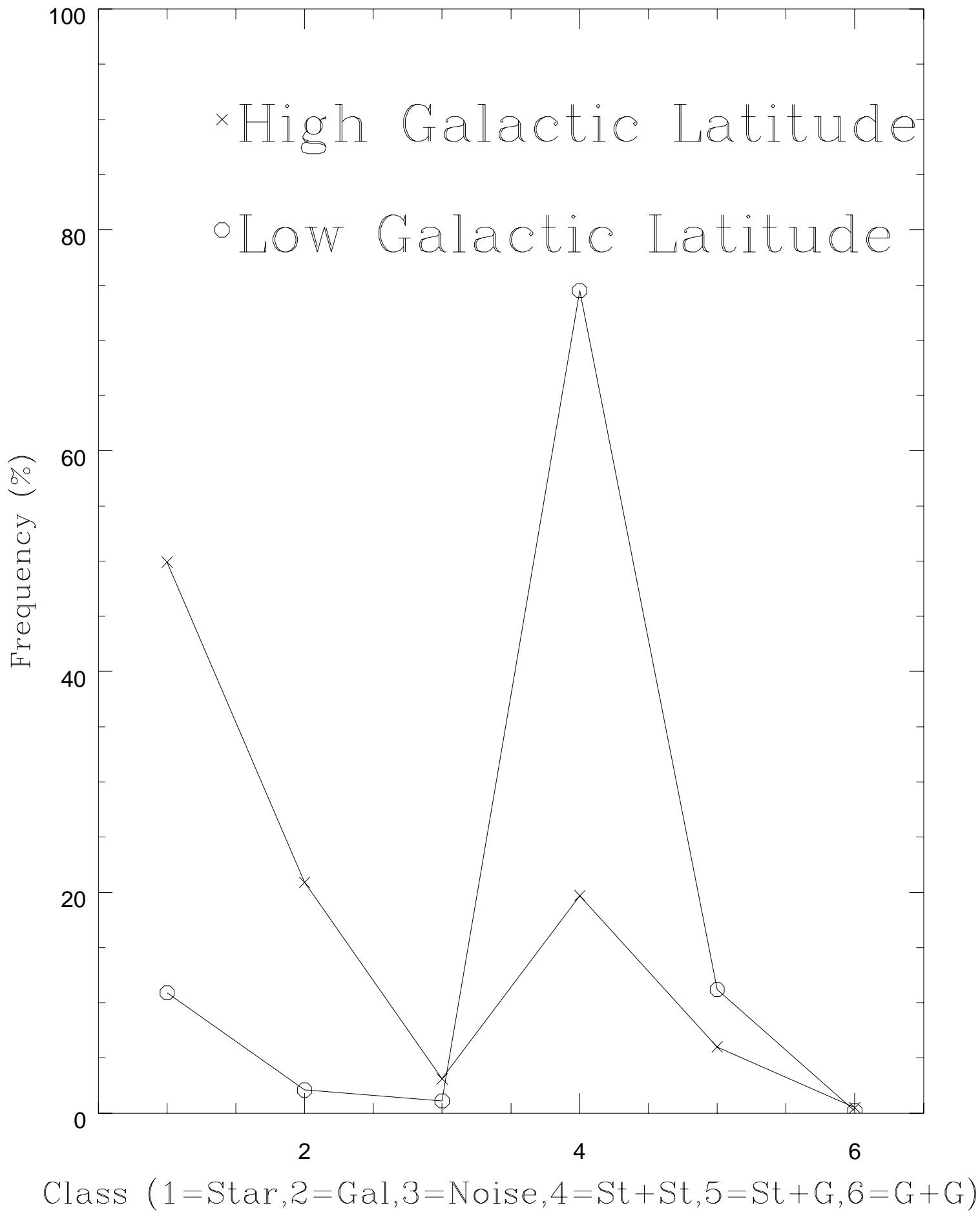
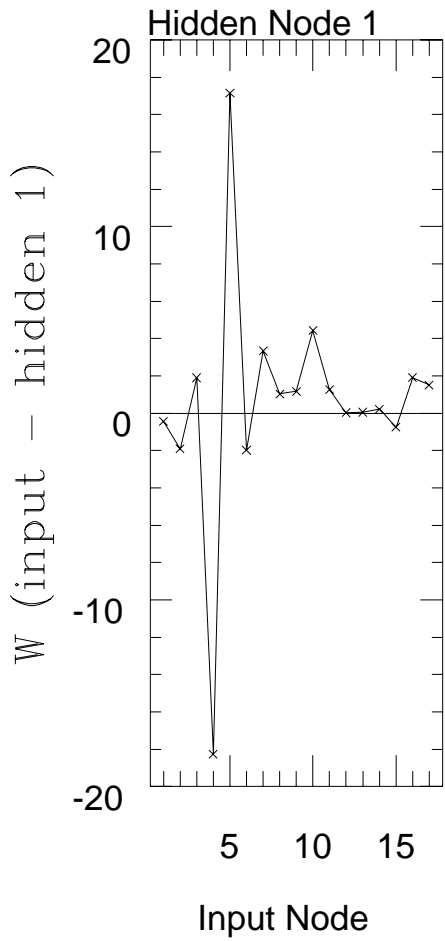
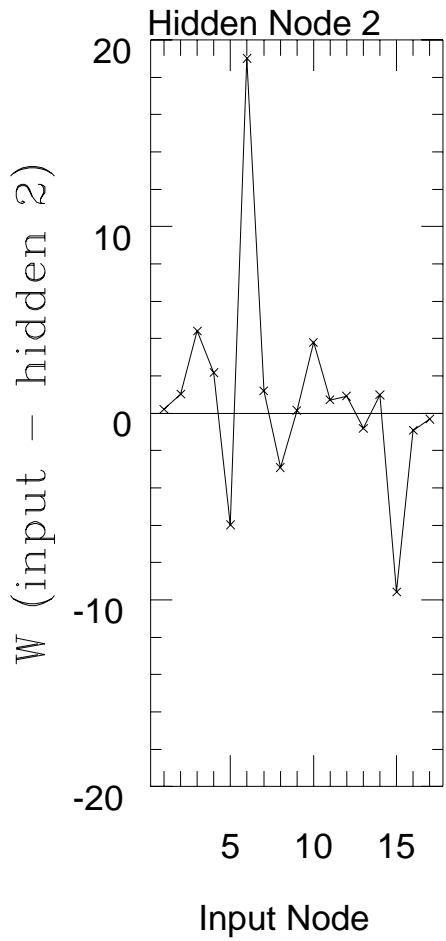
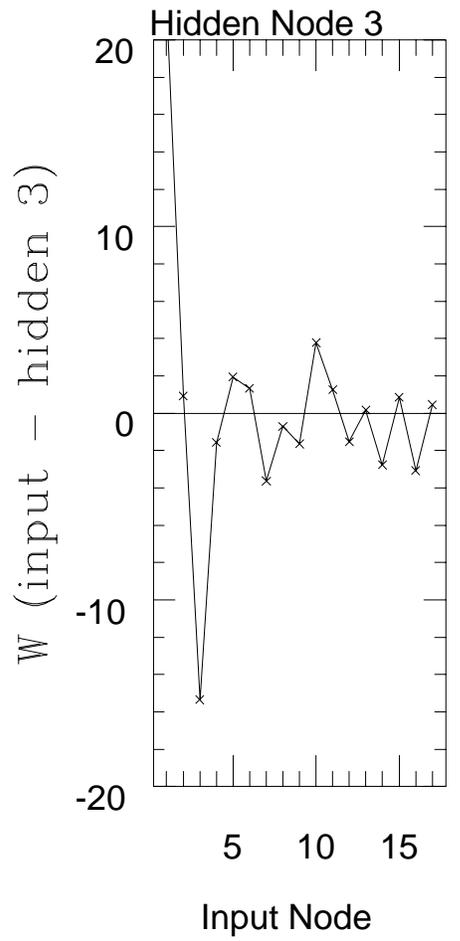
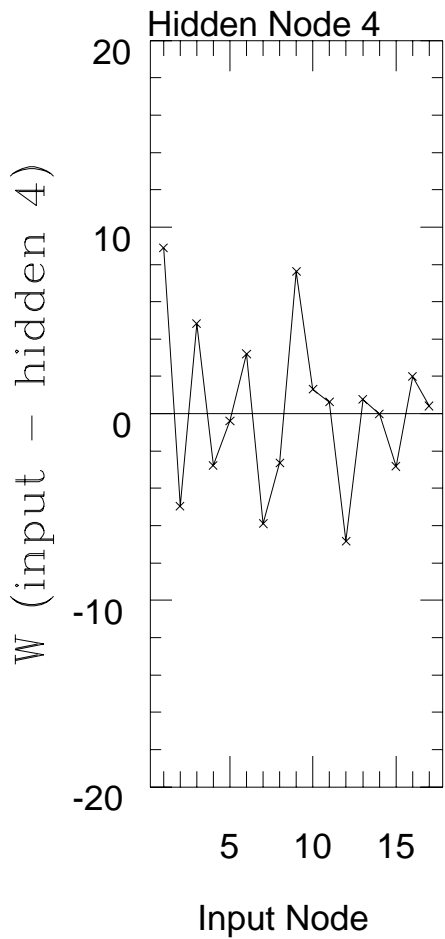
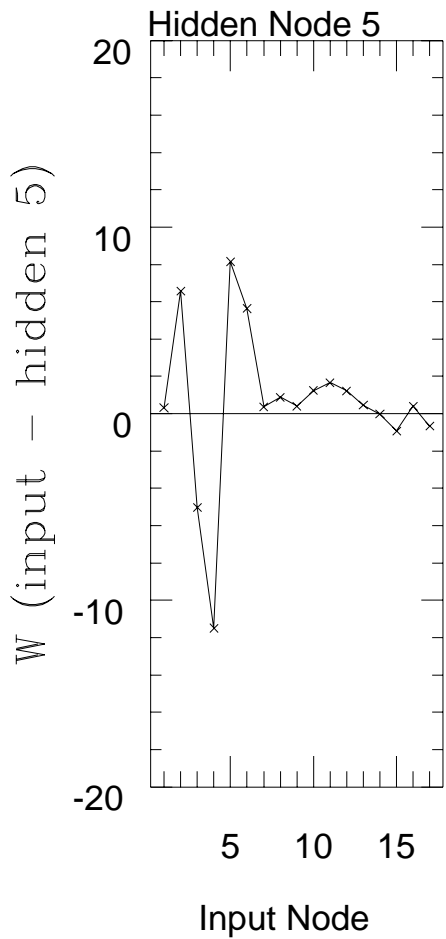
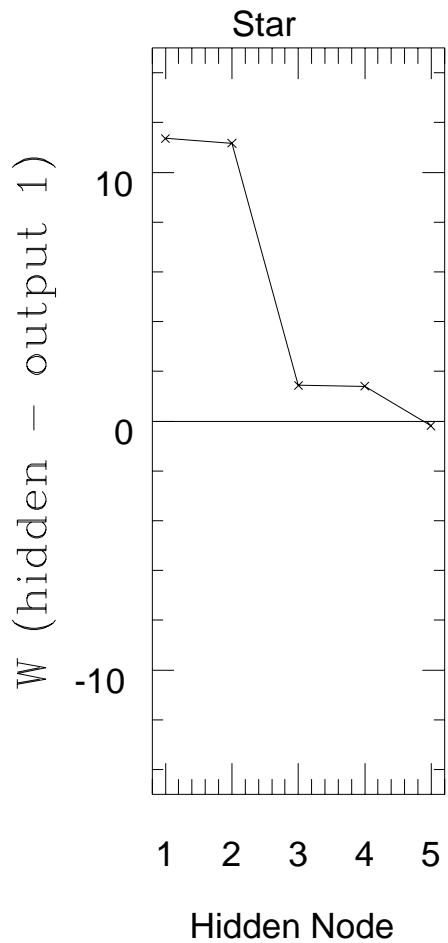
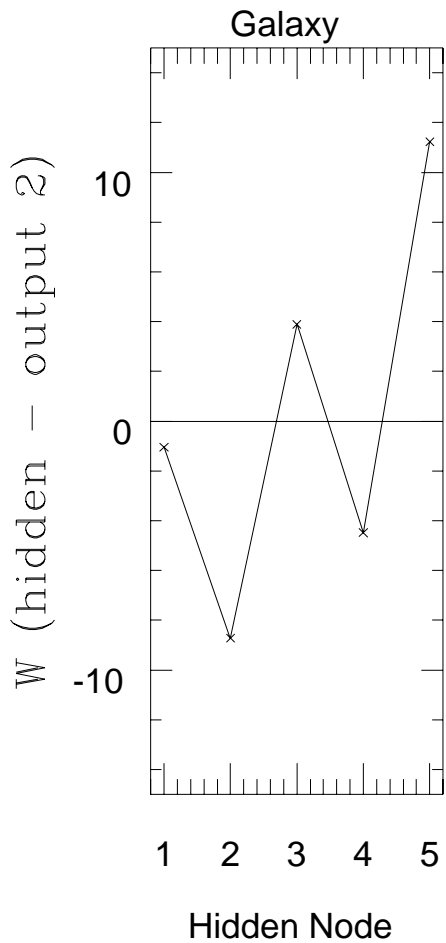
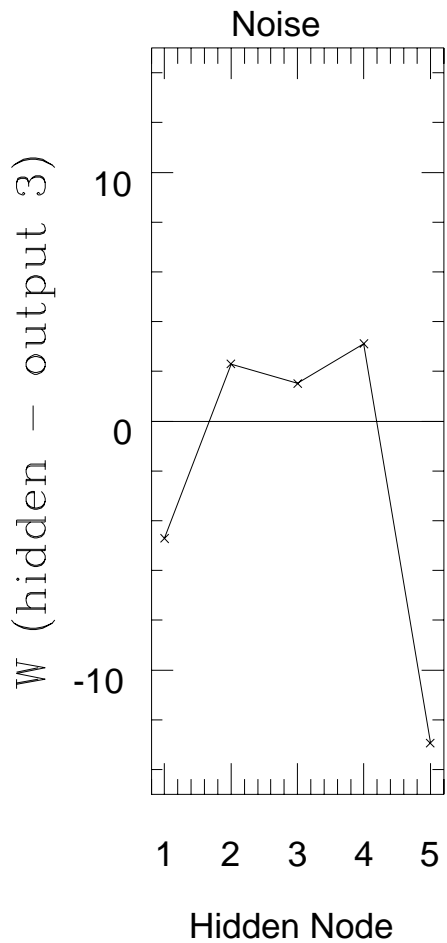
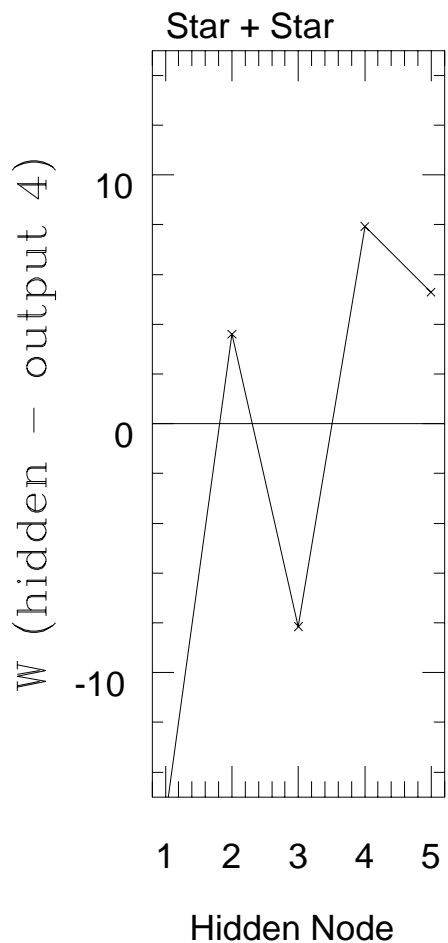
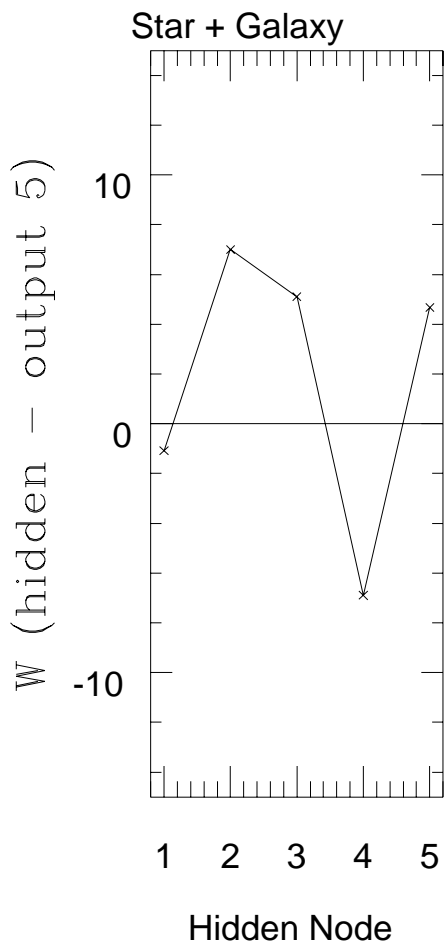
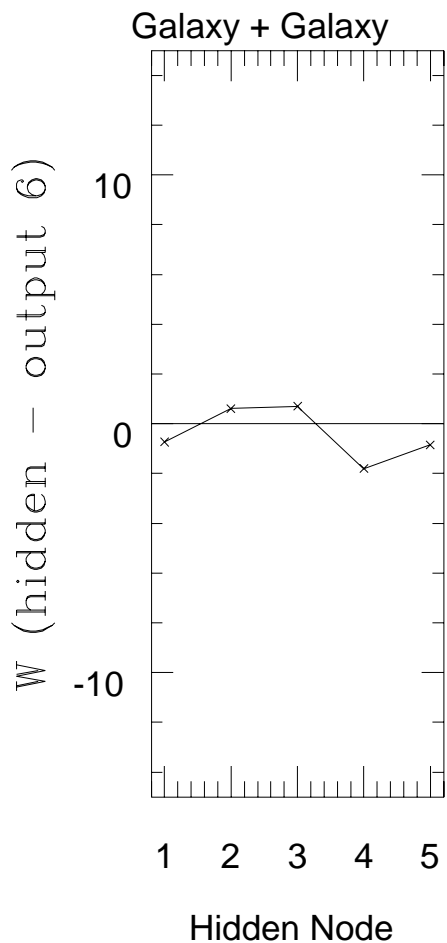
\end{document}